# Growth and inelastic light scattering studies on $Sr_2Nb_2O_7$ single crystals

M. Suganya[1,*], P. Vijayakumar[2], V. Sivasubramanian[2,3], K. Ganesan[2,3], R.M. Sarguna[2], Amirdha Sher Gill[1], S. Ganesamoorthy[2,3]

[1]*Sathyabama Institute of Science and Technology, Chennai-600119, Tamilnadu, India*

[2]*Indira Gandhi Centre for Atomic Research, Kalpakkam, India.*

[3]*Homi Bhabha National Institute, Training School Complex, Anushaktinagar, Mumbai-400094, India*

*\*Corresponding author: <u>suganyatvmalai29@gmail.com</u>*

**ABSTRACT**

A crack free and high structural quality $Sr_2Nb_2O_7$ single crystals were grown by the optical float zone method using optimized growth parameters. Laue pattern confirms single crystalline nature of the grown crystal. Temperature dependent Raman and Brillouin light scattering studies reveal a significant shift in phonon modes across normal to incommensurate phase transition ($T_{n-in}$) which occurs ~ 488 K. In the temperature range from 900 down to 500 K, two optical phonon modes about 63 (B1 mode) and 54 $cm^{-1}$ (A1 mode) were observed. The frequency of A1 mode strongly decreases with an increase in temperature above the $T_{n-in}$ while the frequency of this mode almost remains constant below the $T_{n-in}$. In contrast, the frequency of B1 phonon mode is found to increase with temperature in the range of 500 - 900 K but it does not display a significant shift below the phase transition temperature. In addition, in the in-commensurate phase (T< 488 K), a new optical phonon mode (M1) at ~ 35 $cm^{-1}$ also begins to appear and exhibits strong stiffening behavior with increase in temperature in the range of 300 – 488 K. Moreover, the anomalous behavior of the acoustic phonon across $T_{n-in}$ were further probed using Brilliouin scattering. Longitudinal acoustic phonon mode at 41 GHz exhibits strong change in slope near $T_{n-in}$. In addition, the transverse acoustic modes at 28.6 and 22.4 GHz also exhibit strong anomalies with minimum in frequency near $T_{n-in}$. The inelastic light scattering studies provide valuable information on the phase transition.

## 1. Introduction

Sr$_2$Nb$_2$O$_7$, is a ferroelectric material having layered perovskite orthorhombic structure with high Curie temperature (T$_c$ = 1615 K). Due to its outstanding piezoelectric and ferroelectric characteristics, Sr$_2$Nb$_2$O$_7$ crystals have been widely studied [1–9]. Owing to its interesting features, the research focus has been placed on its usage in cutting-edge technologies including actuators in micro-electro-mechanical systems [8], acoustic tools and non-volatile RAM for its use in capacitors [10]. It is also a suitable choice for green materials from an environmental standpoint because it is lead-free.

Sr$_2$Nb$_2$O$_7$ is one of a group of complicated layered compounds with the chemical formula of Sr$_n$Nb$_n$O$_{3n+2}$. The number of NbO$_6$ octahedra that make up the slab thickness is represented by n (n = 4, for Sr$_2$Nb$_2$O$_7$). At room temperature, the SrNbO$_3$ slabs are arranged along the b-axis of the orthorhombic phase. At high temperature, the NbO$_6$ network distorts and tilts thereby undergoing successive phase transitions. Sr$_2$Nb$_2$O$_7$ undergoes paraelectric (phase I) to ferroelectric phase transition (phase II) at about 1615 K [11]. In phase II, upon transition to the ferroelectric phase the spontaneous polarization appears along *c*-axis and the space group changes from *Cmcm* to *Cmc2$_1$*[8]. Sr$_2$Nb$_2$O$_7$ exhibits a second order phase transition from normal to in-commensurate (T$_i$) phase (phase III) at about 488 K. This phase transition was revealed in dielectric permittivity, elastic compliance coefficient and electro-mechanical coupling factor. At 117 K, another phase transition (phase IV) takes place. The dynamics of phase transition in Sr$_2$Nb$_2$O$_7$ were studied by employing Raman and Brillouin inelastic scattering [1,3,12–16]. In Sr$_2$Nb$_2$O$_7$, a soft mode in the ferroelectric phase (in the phase II) was observed by Raman scattering [12]. The frequency of Raman active phonon mode at 54 cm$^{-1}$ (A1 mode) was found to decrease with an increase in temperature. Below T$_i$ (in phase III) another soft mode at 35 cm$^{-1}$ (M1 mode) was observed suggesting the displacive nature of the in-commensurate phase transition. Further, the phase transition temperature was modulated by doping such as Ba at A-site in Sr$_2$Nb$_2$O$_7$ [16]. In addition, the pressure induced incommensurate to commensurate transition was also observed at 6.5 ± 0.2 GPa in room temperature [3]. No studies have been carried out so far to understand the phonon mode behavior of Sr$_2$Nb$_2$O$_7$ on decreasing the temperature from well above the T$_i$ down to room temperature.

Brillouin light scattering studies in $Sr_2Nb_2O_7$ revealed the very weak anomaly of frequency of longitudinal acoustic (LA) phonon mode at 43 GHz across the in-commensurate phase transition [13]. However, no transverse acoustic (TA) phonon mode was observed in back-scattering geometry. In the present work, we report on the growth of $Sr_2Nb_2O_7$ single crystals using optical floating zone method and study the normal to incommensurate phase transition through phonon anomalies observed from Raman and Brillouin light scattering methods. Importantly, we observe two TA phonon modes at 29 (TA1) and 23 (TA2) GHz in Brillouin scattering and discuss their anomalous behavior across $T_{in-n}$ phase transition.

## 2. Experimental procedure

$Sr_2Nb_2O_7$ was synthesized using starting precursors of $SrCO_3$ and $Nb_2O_5$ of 4N purity. The mixed chemicals were sintered at a temperature of 1573 K for 24 h with intermediate regrinding. The powder X-ray diffraction (PXRD) was carried out using STADI MP diffractometer for confirming its phase formation. Fig. 1 shows the PXRD pattern of the synthesized polycrystalline powders and it confirms the formation of single phase $Sr_2Nb_2O_7$ with perovskite structure. Rietveld refinement of the XRD pattern was performed to obtain the crystallographic parameters. The refined pattern is shown in Fig. 1. The lattice parameters were obtained to be $a = 3.9559(3)$ Å, $b = 26.7927(4)$ Å and $c = 5.7042(3)$ Å which are very close to the reported value for the crystal [17].

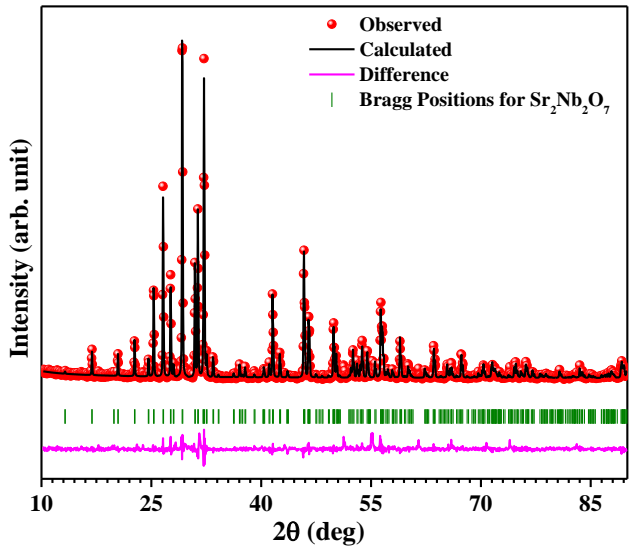

Fig. 1. Powder X-ray diffraction pattern of synthesized polycrystalline $Sr_2Nb_2O_7$

After the phase confirmation, the polycrystalline feed and seed rods of required length and diameter were prepared under the pressure of 70 MPa. The growth was initiated with polycrystalline seed and feed rods using optical floating zone technique and the growth rate was kept at 8 mm/h. Counter clock rotation of feed and seed rods at 30 rpm speed was employed. The growth ambiance was air at the flow rate of 1 *l/min*. During the growth, bubble formation was continuously originated in the melt region and sometimes they coalesce and even burst out thereby disturbing the solid-liquid interface. Due to high anisotropic nature of $Sr_2Nb_2O_7$, the grown crystal contained a lot of cracks. By employing cleaved crystal as seed, a large crack free crystals of reasonable dimension were obtained. One such grown crystal which was cleaved is shown in Fig 2a. A large bubble shown in the marked region was retained in the grown crystal. The crystal wafer was cut from the crack free region and polished with diamond paste (Fig 2b). The polished wafer was subjected to structural and optical studies.

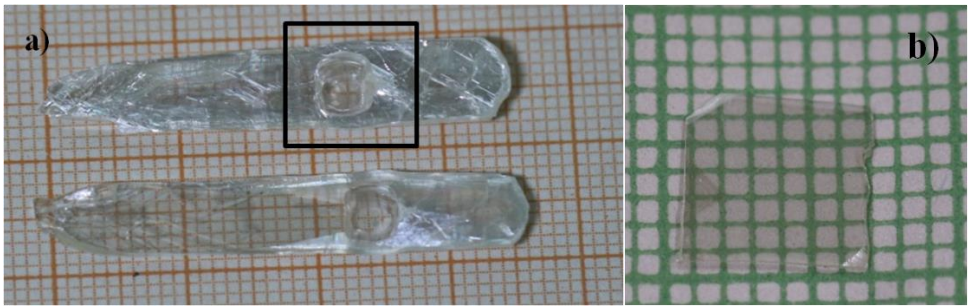

Fig. 2 (a) Photograph of $Sr_2Nb_2O_7$ single crystal that was cleaved along growth axis into two parts. The square marked area indicates the bubble formation during crystal growth. (b) Photograph of polished wafer of $Sr_2Nb_2O_7$ single crystal

The Raman measurements as a function of temperature were carried out in back scattering geometry in the frequency range of 10 – 100 $cm^{-1}$ using Jobin Yvon U1000 Raman spectrometer. Micro-Brillouin light scattering measurements were performed as a function of temperature using 3+3 tandem Fabry-Perot interferometer in the free spectral ranges (FSR) of 75 GHz [18]. For both Raman and Brillouin scattering studies, diode pumped solid state laser with wavelength of 532 nm was used as excitation source. The temperature dependent Raman and Brillouin scattering studies in range 870–300 K was performed using a Linkam (THMS 600) stage. The Raman frequency of phonon modes were deduced by fitting the experimental spectra with the Lorentzian function. Brillouin scattering spectra were fitted with Voigt function to extract the phonon frequency. For Brillouin scattering, the Gaussion part of the Voigt function was kept constant to account for the instrumental broadening.

## 3. Results and discussion

### 3.1 X-Ray diffraction

Figure 3a shows the Laue diffraction pattern which confirmed the single crystallinity of the grown $Sr_2Nb_2O_7$ crystal. Further, X-ray rocking curve analysis was performed on the cleaved (010) wafer and the FWHM of 450 arc second was obtained confirming the reasonable quality of the grown crystal (Fig 3b). Moreover, the PXRD pattern of the powders obtained by grinding the $Sr_2Nb_2O_7$ single crystal matches well with the reported data [JCPDS: 01-070-0114] with orthorhombic structure.

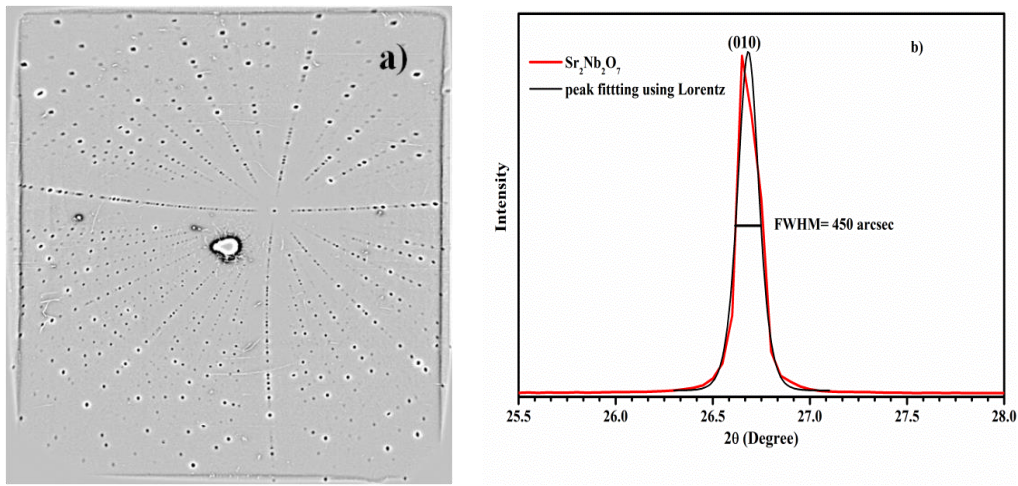

Fig.3. (a) Laue diffraction pattern of $Sr_2Nb_2O_7$ single crystal and (b) X-ray rocking curve on cleaved plate of $Sr_2Nb_2O_7$

### 3.2 Chemical etching

Chemical etching is extensively used by material scientists to study dislocation density in grown single crystals. Chemical etching can create structures on a crystal surface such as etch spirals, etch hillocks, flat-bottomed pits, and so on. The surface feature of the mechanical polished $Sr_2Nb_2O_7$ crystal wafer is shown in Fig 4a. $Sr_2Nb_2O_7$ crystal wafer was etched using $H_2SO_4$ for various etching duration. $H_2SO_4$ etchant resulted in circular and rice husk shape etch patterns on 10 min etched surface as shown in Fig 4b. The net etch pit density is found to be ~ $1.7 \times 10^6$ $cm^{-2}$ which signifies the reasonable structural quality of the grown crystal.

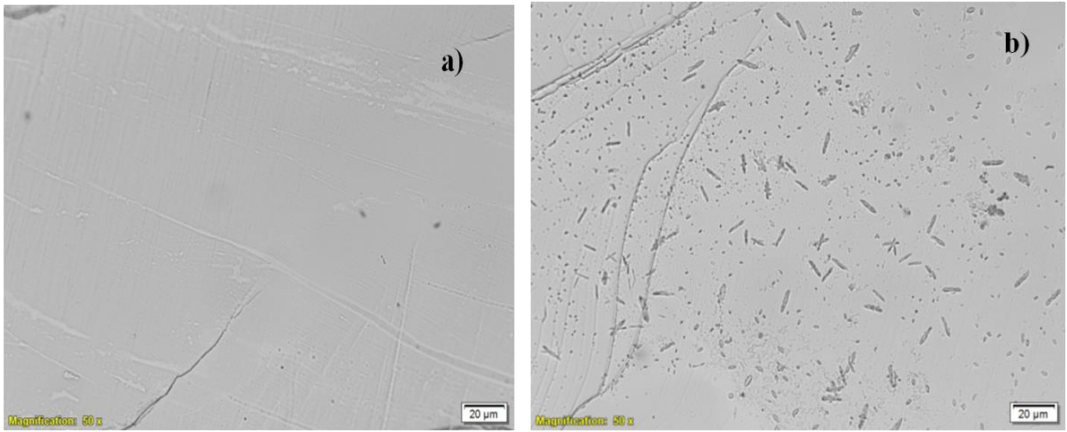

Fig. 4 Optical micrograph of (a) mechanically polished surface and (b) chemically etched surface pattern on $Sr_2Nb_2O_7$ crystal wafer

### 3.3 Raman scattering studies

Fig. 5 depicts the low frequency Raman spectra at some selected temperatures. For $Sr_2Nb_2O_7$ crystal with different crystallographic structure, the number of allowed Raman active phonon modes had been reported using factor group analysis [19]. For room temperature in-commensurate phase (Phase III), the number of zone centre Raman active phonon modes is estimated to be 132 assuming that the $Sr_2Nb_2O_7$ crystal has 4 formulae units in extended double cell structure. Of the 132 modes, there are 44 $A_1$, 22 $A_2$, 22 $B_1$ and 44 $B_2$ vibration modes but three acoustic phonon modes are inactive in Raman spectroscopy [14]. Further, in Phase II and Phase I, there are 66 Raman active modes present that are estimated by assuming the $Sr_2Nb_2O_7$ crystal with 2 formulae units. This study only reports only the Raman modes with low frequency (<80 $cm^{-1}$). Above $T_i$ (488 K), two Raman modes 63 (B1 mode) and 54 (A1 mode) $cm^{-1}$ are observed [14]. These modes are responsible for the paraelectric to ferroelectric and normal to incommensurate phase transitions respectively. Temperature dependence of these modes is shown in Fig 6. The frequency of B1 mode decreases with a decrease in temperature with a shallow maximum at $T_i$.

On the other hand, with the decrease in temperature, the frequency of A1 mode increases and attains a maximum of 54 cm$^{-1}$ at $T_i$. Below $T_i$, the frequency of this mode does not show any noticeable changes. This behavior is not revealed in previously reported Raman data [12]. This Raman mode also did not feature in the previous Raman data which may be due to the difference in symmetry of the measurement configuration [12]. However, this mode is observed in IR spectra and did not exhibit any changes across the phase transition temperature $T_i$ [19]. The increase in the frequency of B1 mode followed by the decrease in the frequency of A1 mode with the decrease in temperature suggests mode repulsion between these two modes. Below $T_i$, a new mode near 35 cm$^{-1}$ (M1 mode) appears as shown in Fig 6. The frequency of M1 mode decreases with decrease in temperature in the in-commensurate phase. The temperature dependence of this mode is consistent with that reported in IR studies [19].

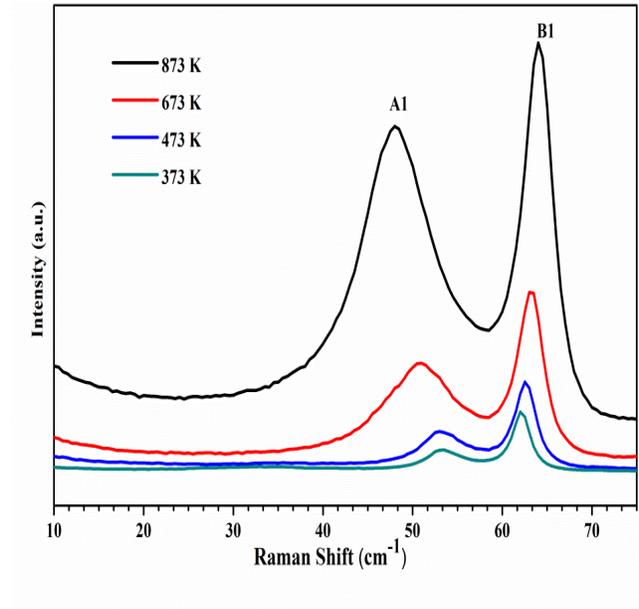

Fig. 5. The variation of Raman spectra of Sr$_2$Nb$_2$O$_7$ crystal at some selected temperatures

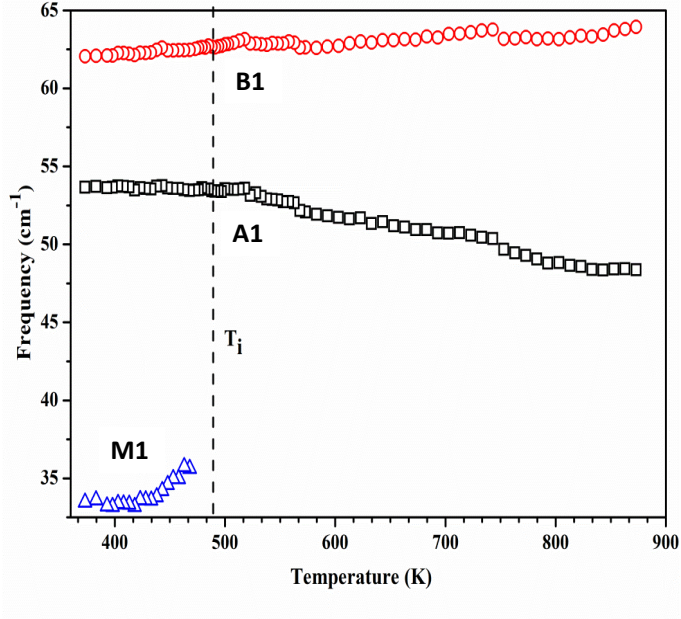

Fig 6. Temperature dependence of frequency of optical phonon modes of $Sr_2Nb_2O_7$

### 3.4 Brillouin scattering studies

To study acoustic phonon anomalies across $T_{n-in}$, Brillouin light scattering studies were carried out in backscattering geometry as a function of temperature. Measurements were carried out on (010) face of the single crystal of $Sr_2Nb_2O_7$ in the $b(c, c+a)\bar{b}$ geometry wherein the incident light is vertically polarized and the scattered light is unpolarized. Brillouin spectra of $Sr_2Nb_2O_7$ at some selected temperatures are shown in Fig 7. A longitudinal acoustic (LA) phonon mode at 41 GHz and two transverse acoustic (TA) modes at 28.6 (TA1) and 22.4 (TA2) GHz has been observed at room temperature. According to the symmetry of the crystal, the elastic constant tensor ($C_{ij}$) of the (010) wafer is calculated from the LA and TA phonons using the equation [20],

$$C_{ij} = \rho (V_{ij})^2, \quad V_{ij} = \frac{\lambda_0 \upsilon}{2n} \quad \text{............ (1)}$$

where, $V_{ij}$ and $\upsilon$ are the velocity and frequency of the LA or TA phonons; $\rho$ and *n* are the mass density and refractive index of the crystal and it is taken as 5.17 g/cm$^3$ and 2.05 ± 0.02, respectively [13]. The $\lambda_0$ is the wavelength of the incident laser, 532 nm. For the orthorhombic crystal structure of $Sr_2Nb_2O_7$, the independent elastic constant tensor $C_{ij}$ reduces to nine components viz. $C_{11}$, $C_{22}$, $C_{33}$, $C_{12}$, $C_{13}$, $C_{23}$, $C_{44}$, $C_{55}$, and $C_{66}$ in the Voigt notation [21]. Here, the $C_{11}$, $C_{22}$ and $C_{33}$ are elastic stiffness constant along the three unit-cell crystal axis. The other constants, $C_{44}$, $C_{55}$ and $C_{66}$, are associated with shear phonon modes, of which the stiffness constant $C_{66}$ is associated with TA1 phonon mode and the constants $C_{44}$ and $C_{55}$ are equal for the space group $Cmc2_1$ and it is associated with TA2 phonon mode [21]. Using the Equation (1), the room temperature elastic constant $C_{22}$ is estimated to be ~ 160 GPa and it is consistent with the reported value [13]. Further, the stiffness constants $C_{44}$ (= $C_{55}$) and $C_{66}$ are calculated to be about 43.7 and 71.1 GPa respectively, at room temperature. Moreover, the variation of $C_{22}$, $C_{44}$ and $C_{66}$ in the temperature range from 400 to 900 K is shown Fig. 8 along with their respective mode LA, TA1 and TA2 frequency. Here, for calculating elastic constant at high temperatures, the refractive index is assumed to be independent of temperature.

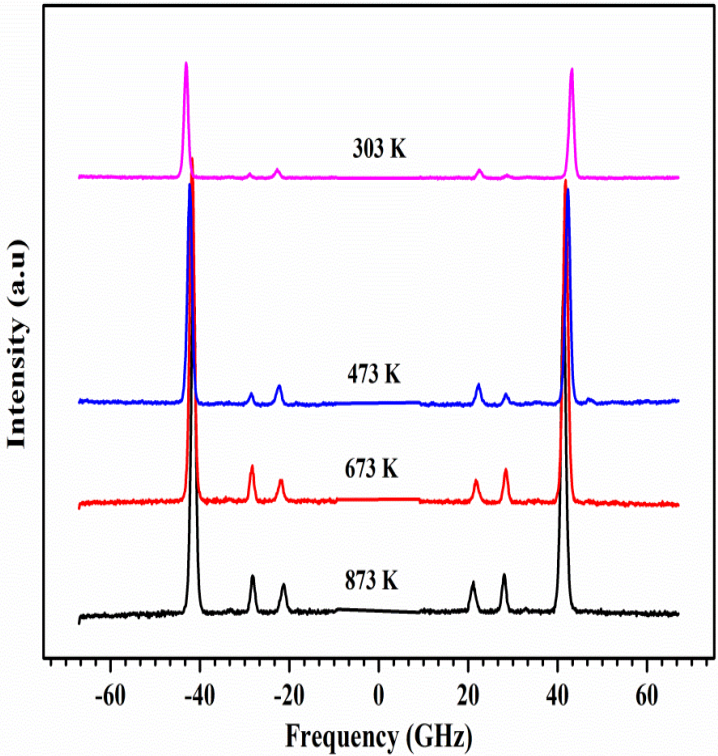

Fig. 7. Brillouin spectra of $Sr_2Nb_2O_7$ crystal at some selected temperatures

The temperature variation in the frequency of LA, TA1 and TA2 phonon modes as shown in Fig. 8 do show a shallow slope change at $T_i$ (488 K) consistent with the reported data [13]. TA1 phonon mode exhibits a sharp decrease in frequency and that of TA2 exhibits a marginal decrease near $T_i$. The softening of TA1 near $T_i$ was not reported in the previous study. Since TA2 exhibits only marginal changes, its features will not be discussed further. In ferroelectric materials, the most likely origin of the softening of the acoustic phonon near the phase transition region is due to the coupling between the polarization and strain. Nature of coupling between the acoustic phonon and polarization enables one to distinguish between the relaxation and the fluctuation mechanisms, about the anomalous changes in frequency of acoustic phonon [22].

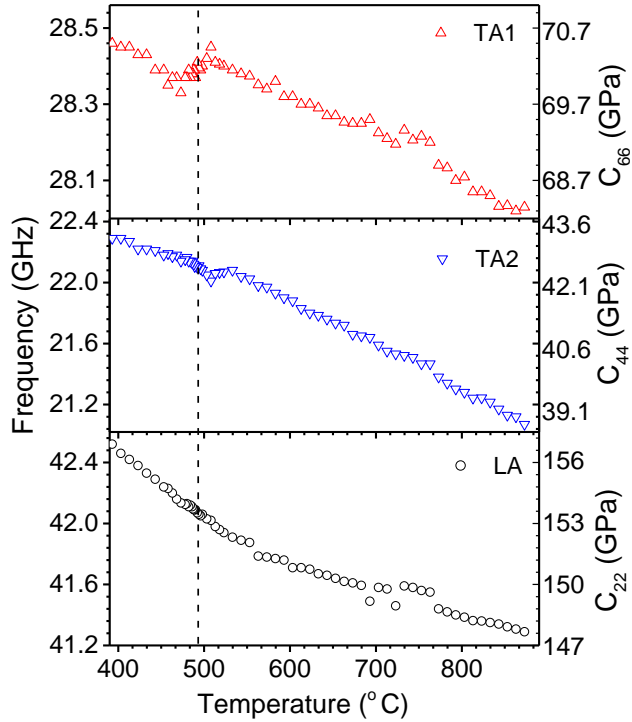

Fig. 8 Temperature dependence of frequency and elastic stiffness constant corresponding to LA, TA1 and TA2 modes of $Sr_2Nb_2O_7$.

The relaxation mechanism crop up due to the linear coupling of strain with polarization. Its manifestation is observed only in the ordered phase below $T_c$ where $P_s \neq 0$. For linear coupling, the acoustic phonon softens as $|T-T_c|^{-1}$ on both sides of the transition. Due to the quadratic coupling of polarization with strain, a fluctuation mechanism arises which leads to an abrupt change in the frequency of the acoustic phonon at $T_c$. As can be seen from Fig 8, the frequency of TA1 mode increases linearly and abruptly decrease near $T_i$ and once again show a linear increase. The linear increase above and below $T_i$ is generally attributed to the normal an-harmonic behavior of the phonon mode. Hence the abrupt change in the frequency of TA1 near $T_i$ is due to the quadratic coupling of polarization with strain. However, the relaxation time of TA1 could not be calculated due to the large scattering in the value of width near the transition region.

## 4. Conclusion

Growth conditions were optimized to yield crack free $Sr_2Nb_2O_7$ single crystals. The quality and orientation of the crystals were confirmed by Laue diffraction. Hardening of the optical phonon mode at 54 cm$^{-1}$ followed by the softening of the phonon mode at 66 cm$^{-1}$ towards the in-commensurate phase transition temperature $T_i$ was observed through Raman studies. Below $T_i$, 54 cm$^{-1}$ mode becomes temperature independent, while that of the mode at 66 cm$^{-1}$ exhibit a softening behaviour below $T_i$. A new optical phonon mode at 35 cm$^{-1}$ was observed in the in-commensurate phase, which continues to show a softening behavior with decrease in temperature. Brillouin light scattering studies revealed the softening of transverse acoustic phonon mode at 28.6 GHz near $T_i$. The softening of this mode is suggested being due to the quadratic coupling between the polarization and strain.

**Declaration of interests**

The authors declare that they have no known competing financial interests or personal relationships that could have appeared to influence the work reported in this paper.

**Acknowledgment:**

The authors M.S and A.S.G gratefully acknowledge University Grants Commission (UGC) Indore, India under UGC-DAE-CSR (CSR-KN/CRS-87/2016-17/1128) project scheme for financial assistance.

**Data Availability Statement:**

The data that support the findings of this study are available from the corresponding author upon reasonable request.